\documentclass[twocolumn]{aastex63}

\usepackage{graphicx}	
\usepackage{color}
\usepackage{amsmath}	
\usepackage{amssymb}	
\usepackage{mathrsfs}
\usepackage{mathrsfs}
\usepackage{multirow}

\newcommand{\msun}{M_\odot}

\newcommand{\zsun}{Z_\odot}

\newcommand{\cc}{{\rm cm}^{-3}}

\newcommand{\msunyr}{M_\odot~{\rm yr}^{-1}}

\newcommand{\mpc}{{\rm Mpc}}

\newcommand{\pc}{{\rm pc}}
\newcommand{\mum}{\mu {\rm m}}
\newcommand{\kms}{{\rm km~s}^{-1}}

\newcommand{\K}{{\rm K}}
\newcommand{\beq}{\begin{equation}}
\newcommand{\eeq}{\end{equation}}

\shorttitle{Balmer-break and absorption on AGN spectra}
\shortauthors{K.~Inayoshi \& R.~Maiolino}

\begin{document}

\title{
Extremely Dense Gas around Little Red Dots and High-redshift AGNs: \\
A Non-stellar Origin of the Balmer Break and Absorption Features
}

\correspondingauthor{Kohei Inayoshi}
\email{inayoshi@pku.edu.cn}

\author[0000-0001-9840-4959]{Kohei Inayoshi}
\affiliation{Kavli Institute for Astronomy and Astrophysics, Peking University, Beijing 100871, China}

\author[0000-0002-4985-3819]{Roberto Maiolino}
\affiliation{Kavli Institute for Cosmology, University of Cambridge, Madingley Road, Cambridge, CB3 OHA, UK}
\affiliation{Cavendish Laboratory - Astrophysics Group, University of Cambridge, 19 JJ Thomson Avenue, Cambridge, CB3 OHE, UK}
\affiliation{Department of Physics and Astronomy, University College London, Gower Street, London WC1E 6BT, UK}

\begin{abstract}
The James Webb Space Telescope (JWST) has uncovered low-luminosity active galactic nuclei (AGNs) at high redshifts of $z\gtrsim 4-7$, 
powered by accreting black holes (BHs) with masses of $\sim 10^{6-8}~\msun$.
One remarkable distinction of these JWST-identified AGNs, compared to their low-redshift counterparts, is that at least $\sim 20\%$ of them 
present H$\alpha$ and/or H$\beta$ absorption, which must be associated with extremely dense ($\gtrsim 10^9~\cc$) gas in the broad-line region or its immediate surroundings.
These Balmer absorption features unavoidably imply the presence of a Balmer break caused by the same dense gas.
In this Letter, we quantitatively demonstrate that a Balmer break can form in AGN spectra without stellar components, when the accretion disk is
heavily embedded in dense neutral gas clumps with densities of $\sim 10^{9-11}~\cc$, where hydrogen atoms are collisionally excited to the $n=2$ states 
and effectively absorb the AGN continuum at the bluer side of the Balmer limit.
The non-stellar origin of a Balmer break offers a potential solution to the large stellar masses and densities inferred for little red dots (LRDs) 
when assuming that their continuum is primarily due to stellar light.
Our calculations indicate that the observed Balmer absorption blueshifted by a few hundreds $\kms$ suggests 
the presence of dense outflows in the nucleus at rates exceeding the Eddington value.
Other spectral features such as higher equivalent widths of broad H$\alpha$ emission and presence of \ion{O}{1} lines observed in high-redshift AGNs 
including LRDs align with the predicted signatures of a dense super-Eddington accretion disk.
\end{abstract}

\section{introduction}

The James Webb Space Telescope (JWST) is rapidly advancing our exploration of the high-redshift universe. 
With its exceptional sensitivity, JWST has uncovered numerous intermediate/low-luminosity active galactic nuclei (AGNs), 
enabling us to study the representative population of accreting black holes (BHs) at cosmic dawn \citep[e.g.,][]{Onoue_2023,Kocevski_2023,Harikane_2023_agn,Maiolino_2023_JADES,Maiolino_2024a}.

Among the most intriguing discoveries are very compact, red-colored sources with broad-emission line (FWHM $\gtrsim 1500~\kms$) features in their spectra
\citep[e.g.,][]{Labbe_2023,Barro_2024,Matthee_2024,Greene_2024}.
These so-called ``little red dots" (LRDs) are considered to be dust-reddened AGNs at $z\sim 4-8$, with bolometric luminosities of 
$L_{\rm bol}\sim 10^{44-47}~{\rm erg~s}^{-1}$ powered by massive BHs with $10^{7-8}~\msun$, if dust-attenuation correction derived 
from the red continua in the rest-frame optical bands is applied.
Remarkably, the cosmic abundance of these LRDs is one order (or two orders, if the whole AGN population is considered) of magnitude higher than what was expected from previous quasar surveys \citep{Kokorev_2024a,Akins_2024,Kocevski_2024}. 
If all these bright LRDs are AGNs, this would imply that the radiative efficiency is approaching the theoretical limit, 
requiring a potential re-evaluation of current observations or theoretical models \citep{Inayoshi_Ichikawa_2024}.

Despite their significance, the properties of the newly identified LRDs remain puzzling, particularly regarding the origin 
of their characteristic v-shaped spectral energy distribution (SED) in the rest-frame UV-to-optical bands \citep{Labbe_2023,Greene_2024,Wang_2024a}. 
Possible explanations for this feature include contributions from galaxies, AGNs, or a combination of both.
Recently, deep spectroscopic observations have revealed a prominent drop near the Balmer limit in the continuum spectra of some LRDs \citep{Greene_2024,Furtak_2024,Wang_2024a,Wang_2024b,Baggen_2024,Kokorev_2024b}, 
suggesting a potential contribution from the host galaxy stellar light. This finding is crucial for understanding the energy source of LRDs.
If LRDs are powered by dusty starburst galaxies alone, the inferred stellar mass would exceed a few times $10^{10}~\msun$, and in some cases reach up to $\sim 10^{11}~\msun$, which would conflict with the standard structure formation framework if such huge masses were formed at $z\gtrsim 7$ \citep{Wang_2024b,Akins_2024,Inayoshi_Ichikawa_2024}.
Additionally, when combined with the extremely compact sizes, the stellar densities would be so high that velocity dispersions could reach $\gtrsim 1000~\kms$, a phenomenon never observed in local or lower-redshift galaxies (\citealt{Baggen_2024}; also \citealt{Hopkins_2010}).
Alternatively, if the light redward of the Balmer break originates from a non-stellar source (with stellar light dominating only at shorter wavelengths), the inferred stellar mass could be significantly lowered, on the order of $\sim 10^9~\msun$, aligning with structure formation models. 
However, this scenario would still need an AGN contribution at longer wavelengths to explain the continuum with a steep red color and broad Balmer emission lines.
Another interpretation suggests that the UV component of the SED could be influenced by a gray dust attenuation curve, resulting from 
the deficit of small-size dust grains, which might explain the v-shaped SED of LRDs \citep{Li_LRD_2024}.

Moreover, an independent line of evidence comes from the detailed analysis of broad hydrogen emission lines in AGNs observed by JWST, 
not only in LRDs but also in unobscured sources \citep{Matthee_2024,Maiolino_2023_JADES,Kocevski_2024,Juodzbalis_2024,Lin_2024}. 
These observations reveal slightly blueshifted and narrow absorption on the broad Balmer lines ($v\simeq 200~\kms$). 
The detection of H$\alpha$ and H$\beta$ in absorption is remarkable, as the $n=2$ states of atomic hydrogen are very short lived and not metastable.
To make such absorption features visible against the Balmer emission profile, extremely high gas densities are required to populate hydrogen atoms into the $n=2$ states. 
In particular, \citet{Juodzbalis_2024} infer that H$\alpha$ and H$\beta$ absorption must be associated with very dense gas along the line of sight 
with $n_{\rm H}>10^9~\cc$, possibly clouds of the broad line region (BLR) or its immediate surroundings.
In nearby AGNs, Balmer absorption lines are rarely observed with a detection rate of $\approx 0.1\%$. 
However, Balmer absorption has been found in at least $10-20\%$ of broad-line AGNs observed by JWST \citep[see Figure 12 of][]{Lin_2024}. 
Since higher-resolution spectroscopy is required to detect these absorption lines, the fraction of $10-20\%$ is likely a lower limit,
and thus a larger fraction of AGNs are probably buried in dense gas distributed over a wide solid angle.

In this Letter, we demonstrate that a Balmer break feature can form in AGN spectra without stellar components, when the accretion disk is heavily embedded in dense neutral gas clumps 
with densities of $\simeq 10^{9-11}~\cc$, where hydrogen atoms are collisionally excited to the $n=2$ states 
and effectively absorb the AGN continuum at the bluer side of the Balmer limit. 
Under these circumstances, the dense gas clump naturally leads to deep absorption on top of the broad Balmer emission lines as observed in JWST AGNs.
We further discuss the physical mechanism of launching dense outflows imprinted in the blueshifted Balmer absorption, and other spectral features of accreting BHs embedded in dense environments.

\begin{figure*}
\centering
\includegraphics[width=85mm]{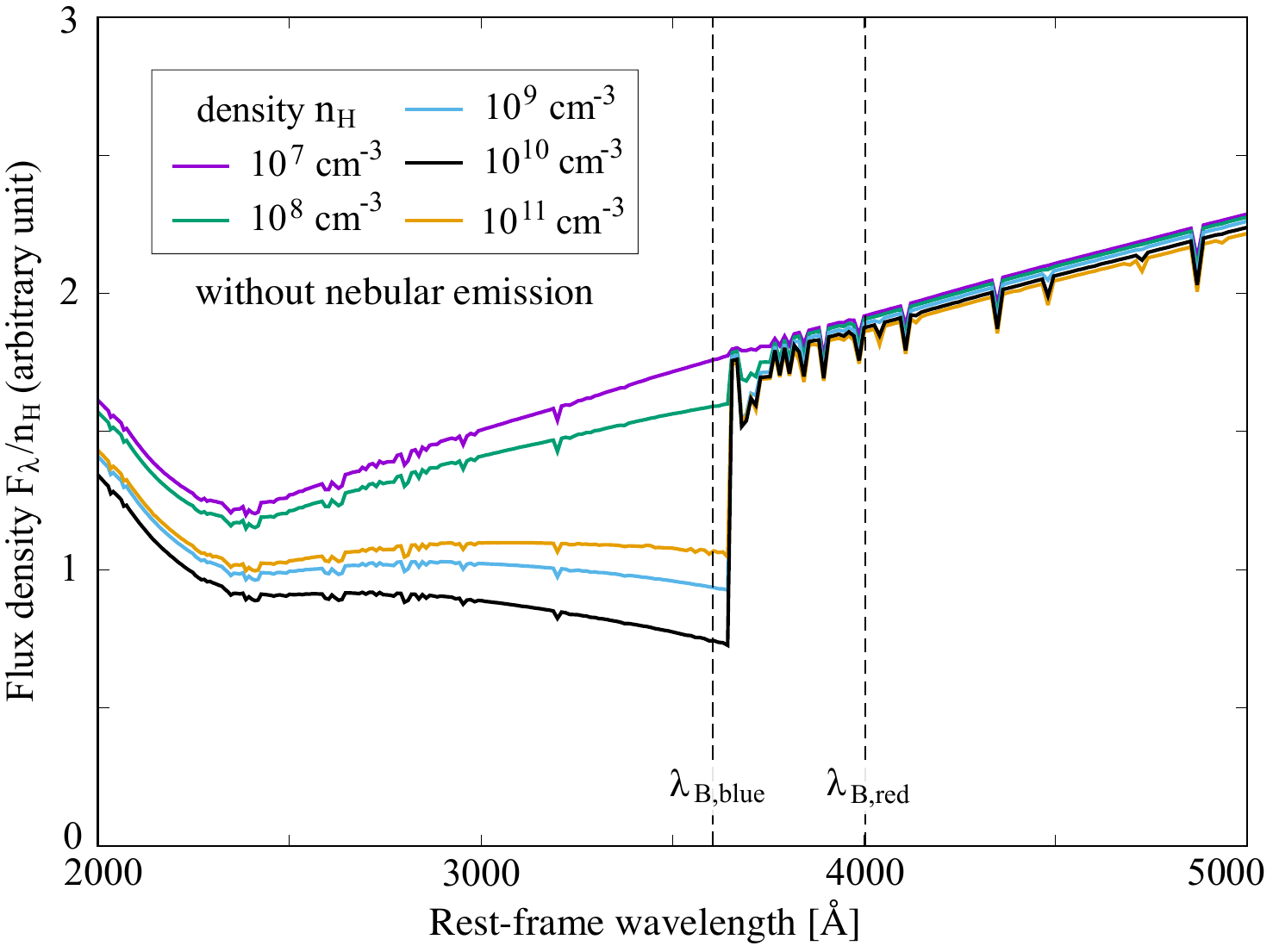}\hspace{5mm}
\includegraphics[width=85mm]{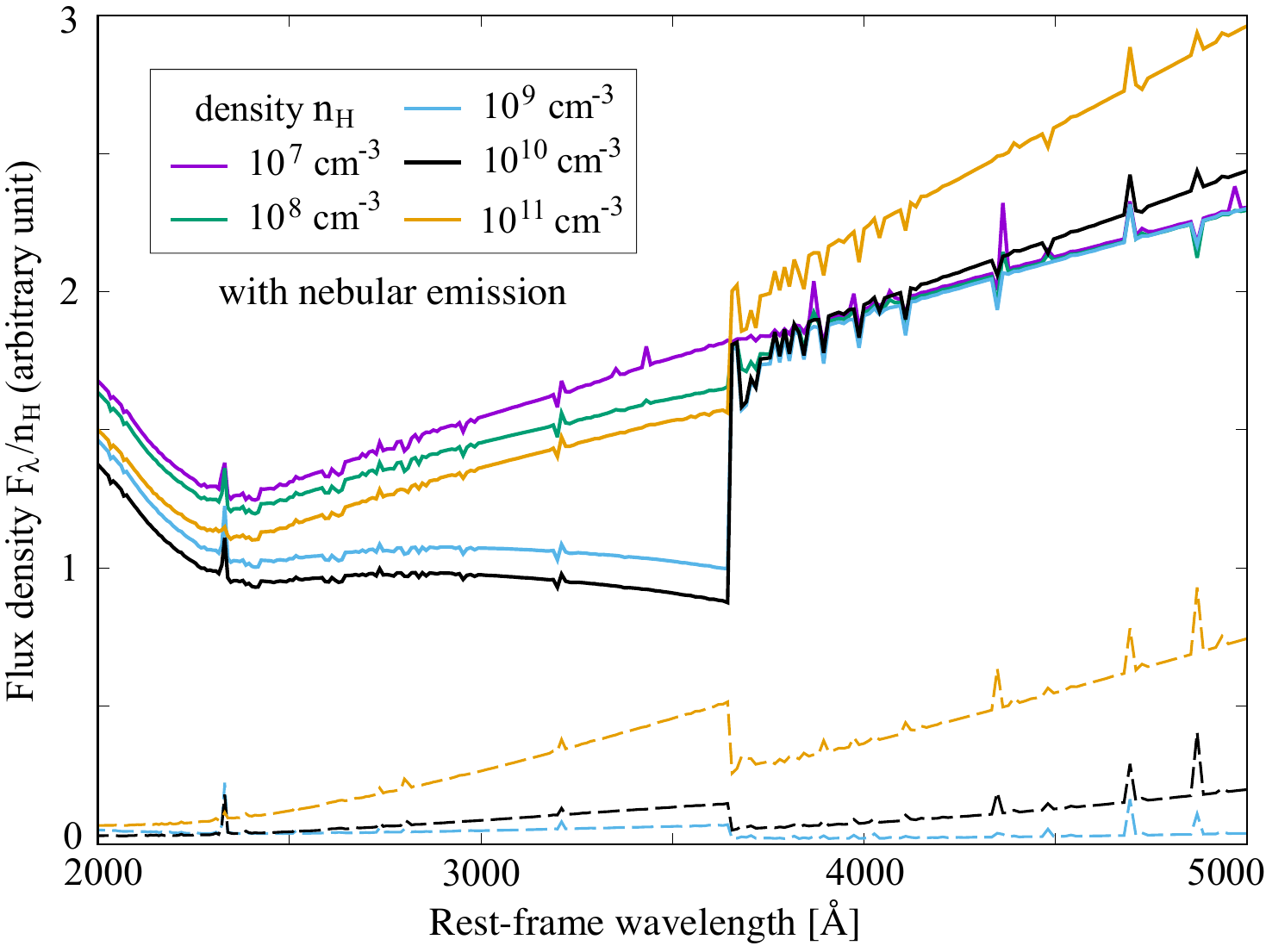}
\caption{Left: AGN SEDs attenuated through a gas slab with a visual extinction of $A_V=3$ mag with $Z=0.1~\zsun$.
Each curve represents the case with different density ($10^7\leq n_{\rm H}/\cc \leq 10^{11}$) and thickness.
With high densities of $n_{\rm H}\simeq 10^{9-11}~\cc$, the SEDs show a deep Balmer break at $\lambda_{\rm B,lim} = 3646~{\rm \AA}$.
Two vertical lines indicate the wavelengths ($\lambda_{\rm B,blue}=3600~{\rm \AA}$ and $\lambda_{\rm B,red}=4000~{\rm \AA}$) used to quantify 
the Balmer-break strength.
Right: Total AGN SEDs including the nebular emission with a covering fraction of $C=0.5$.
For the cases with $n_{\rm H}=10^{9-11}~\cc$, the nebular components are shown separately (dashed).
The Balmer jump feature of the nebular spectrum weakens the Balmer break strength in the total SED
when dense absorbers with $n_{\rm H}\gtrsim 10^{11}~\cc$ surround the AGN with a high covering fraction ($C\gtrsim 0.5$).
}
\label{fig:break}
\vspace{5mm}
\end{figure*}

\section{Balmer break}\label{sec:break}

To quantify the SED shape of an attenuated incident flux from the galactic nucleus, we make use of CLOUDY \citep[C17,][]{Ferland_2017} to perform line transfer calculations 
along with hydrogen level population modeling simultaneously.
In our model, the incident radiation source is an AGN (an accretion disk and non-thermal radiation)
and its spectral shape is assumed to be 
\begin{equation}
    f_\nu \propto {\rm max}\left[ \nu^{\alpha_{\rm uv}}e^{-h\nu/k_{\rm B}T_{\rm bb}},~ r_{\rm x}\nu^{\alpha_{\rm x}}\right],
\end{equation}
where we set the temperature of the big blue bump to $T_{\rm bb}=10^5~\K$\footnote{The characteristic temperature corresponds to the value measured at $r\sim 10~r_{\rm g}$ in an accretion disk around a BH with $M_\bullet = 10^{7-8}~\msun$ accreting at the Eddington rate, where $r_{\rm g}$ is the Schwarzschild radius.
Note that the surface temperature profile saturates within $10~r_{\rm g}$ and declines toward the inner-most stable circular orbit, where the torque-free boundary conditions are imposed \citep{Novikov_Thorne_1973}.
},
the UV and X-ray spectral indices to $\alpha_{\rm uv}=-0.5$ and $\alpha_{\rm x}=-1.5$, 
and the normalization of $r_{\rm x}$ is adjusted so that the spectral slope between $2500~{\rm \AA}$ and $2$~keV becomes $\alpha_{\rm ox}=-1.5$.
The value of $\alpha_{\rm uv}=-0.5$ is consistent with that of the low-redshift composite quasar SED \citep{VandenBerk_2001}.
The X-ray spectral index would be steeper as observed in bright quasars at high redshifts ($\alpha_{\rm x}\lesssim -2$; \citealt{Zappacosta_2023}), however our results are unaffected by the specific value of the index.
The flux density normalization is determined such that the ionization parameter, $U\equiv \Phi_0/(n_{\rm H}c)$, falls within $-2\leq \log U \leq -1$, where $\Phi_0$ is the ionizing photon number flux, $n_{\rm H}$ is the number density of hydrogen nuclei, and $c$ is the speed of light.
In this study, we adopt $\log U=-1.5$ as the fiducial choice.
Note that the Balmer-break strength varies by $\simeq 10-20\%$ depending on the ionization parameter within the range.
The distance of the gas absorber derived using $\log U=-1.5$ is consistent with the cloud kinematics, as discussed in Section~\ref{sec:discussion1}.
We consider a plane-parallel geometry of the absorber assuming that individual clouds have a small cross section.
Then, the total SED is calculated by combining the transmitted and nebular components, with the nebular contribution scaled by a covering fraction $C$ 
for gas absorbers within the hemisphere facing the observer.
For the fiducial model, we assume a visual extinction of $A_V=3$ mag to match the redness observed in the rest-optical continua for LRDs \citep{Matthee_2024,Greene_2024}. 
With a metallicity of $0.1~\zsun$, the column density is approximated to $N_{\rm H}\simeq 5.4\times 10^{22}~{\rm cm}^{-2}(Z/0.1~\zsun)^{-1}$.
We vary the gas density over a broad range of $10^7\leq n_{\rm H}/\cc \leq 10^{11}$, adjusting the slab thickness ($\Delta s$) to maintain a fixed visual extinction.
In this analysis, we do not account for the effects of microscopic turbulence, whose influence on the SED shape will be studied in a forthcoming paper \citep{Ji_2025}.

The left panel of Figure~\ref{fig:break} presents the SEDs of an attenuated AGN for various densities of $10^7\leq n_{\rm H}/\cc \leq 10^{11}$.
When the density is $n_{\rm H}=10^7~\cc$ or lower, the SED maintains a smooth red continuum at $\lambda \gtrsim 2500~{\rm \AA}$ and a slightly blue one at the shorter wavelengths.
The bending of the SED at $\lambda \simeq 2500~{\rm \AA}$ is attributed to the gray dust attenuation curve, though variations in the curve model can lead to different results \citep{Li_LRD_2024}, which we do not explore in detail here. 
As the density increases to $n_{\rm H}=10^8~\cc$, the continuum component at wavelengths shorter than the Balmer limit ($\lambda_{\rm B,lim}=3646~{\rm \AA}$) 
becomes steeper, and the discontinuity across the Balmer limit reaches its peak around $n_{\rm H}=10^{9-10}~\cc$. 
However, as the density further increases to $\simeq 10^{11}~\cc$, the discontinuity weakens but converges in the higher density regime.

\begin{figure}
\centering
\includegraphics[width=83mm]{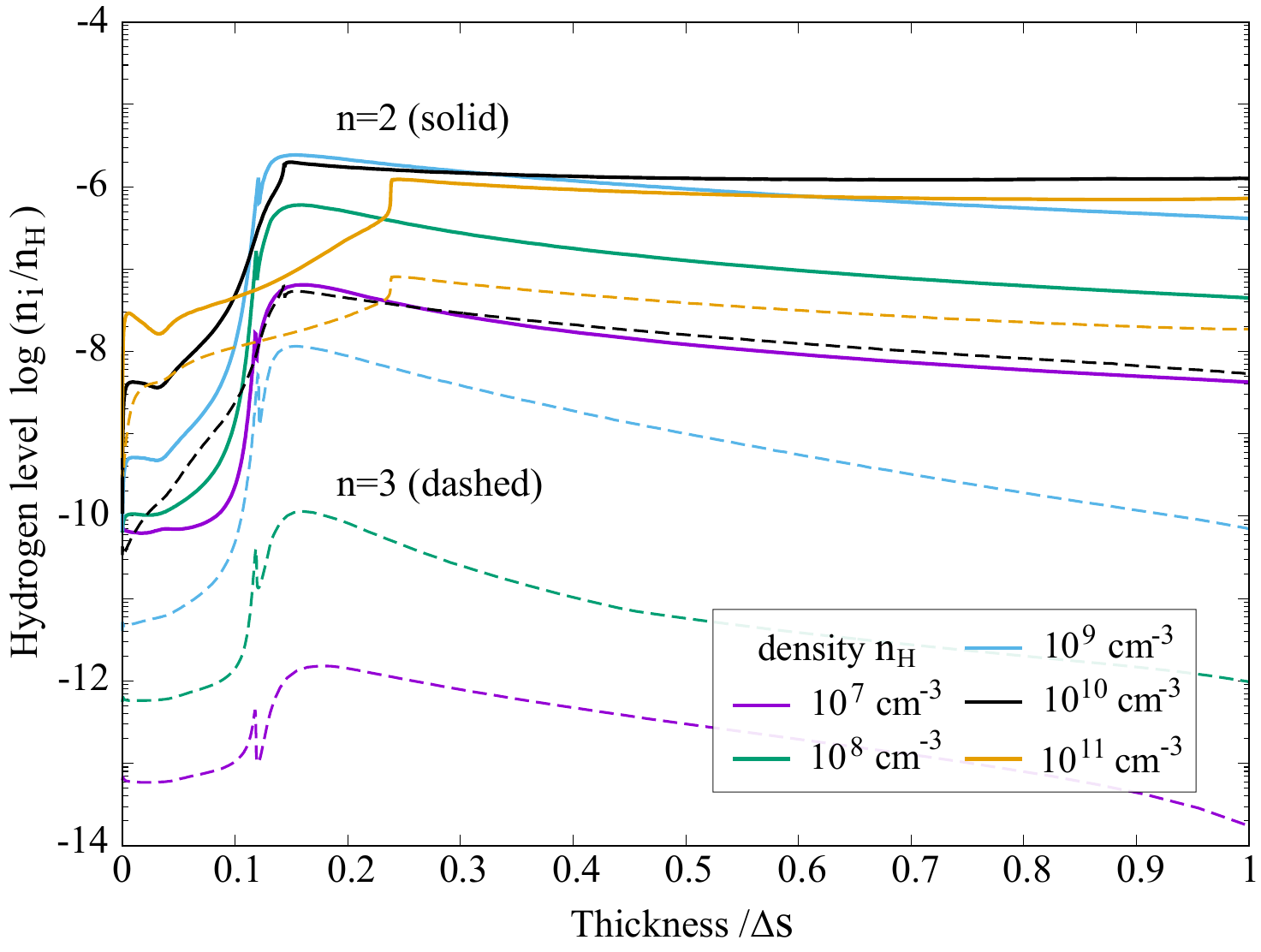}
\caption{
Profiles of the hydrogen level populations in the $n=2$ (solid) and $n=3$ (dashed) states as a function of slab thickness normalized by the total value $\Delta s$ for each density case; $n_{\rm H}=10^7-10^{11}~\cc$.
The values are normalized by the total density of hydrogen nuclei ($n_{\rm H}$), including both neutral and ionized states.
As the density increases, the hydrogen is excited to the higher energy states.
The ratio of $n=2$ states reaches $n_2/n_{\rm H}\simeq 10^{-6}$, which is the equilibrium value with $T\simeq 8000~\K$ through particle collisions. 
}
\label{fig:level}
\vspace{3mm}
\end{figure}

The right panel of Figure~\ref{fig:break} shows the total AGN SEDs (solid) including the nebular emission with a covering fraction of $C=0.5$.
For demonstration, the cases of $n_{\rm H}=10^{9-11}~\cc$ are presented with the nebular components separately (dashed).
For $n_{\rm H}=10^{9-10}~\cc$, the Balmer break imprinted by dense gas absorbers remains prominent.
However, at higher densities ($n_{\rm H}\gtrsim 10^{11}~\cc$) and a high covering fraction ($C\gtrsim 0.5$), 
the Balmer jump feature in the nebular emission reduces the apparent strength of the Balmer break in the total SED.

Figure~\ref{fig:level} shows the number-density ratio of atomic hydrogen in excited states ($n=2$ with solid and $n=3$ with dashed curves) 
to the total density of hydrogen nuclei including neutral and ionized states, as a function of slab thickness. 
At low densities ($n_{\rm H}\lesssim 10^8~\cc$), the number ratios of excited states in both $n=2$ and $n=3$ increase 
proportionally with the density, indicating that atomic hydrogen has not reached a local thermodynamic equilibrium state for a given temperature.
However, as the density reaches $n_{\rm H}\simeq 10^9~\cc$, the ratio of $n=2$ states begins to saturate, approaching the collisional 
equilibrium value of $n_2/n_{\rm H}\simeq 10^{-6}$ for $T\simeq 8000~\K$. 
When the density further increases to $n_{\rm H}=10^{10}~\cc$, the ratio of the $n=3$ state atomic hydrogen also starts to saturate.
The Balmer break strength is closely linked to the population of hydrogen in the $n=2$ state. 
Effective collisional pumping to the $n=2$ state leads to significant attenuation of AGN flux at wavelengths just blueward of the Balmer limit,
naturally producing a Balmer break.
This process is analogous to the Lyman break, which results from neutral atomic hydrogen in the ground state ($n=1$) absorbing ionizing radiation.

\begin{figure}
\centering
\includegraphics[width=85mm]{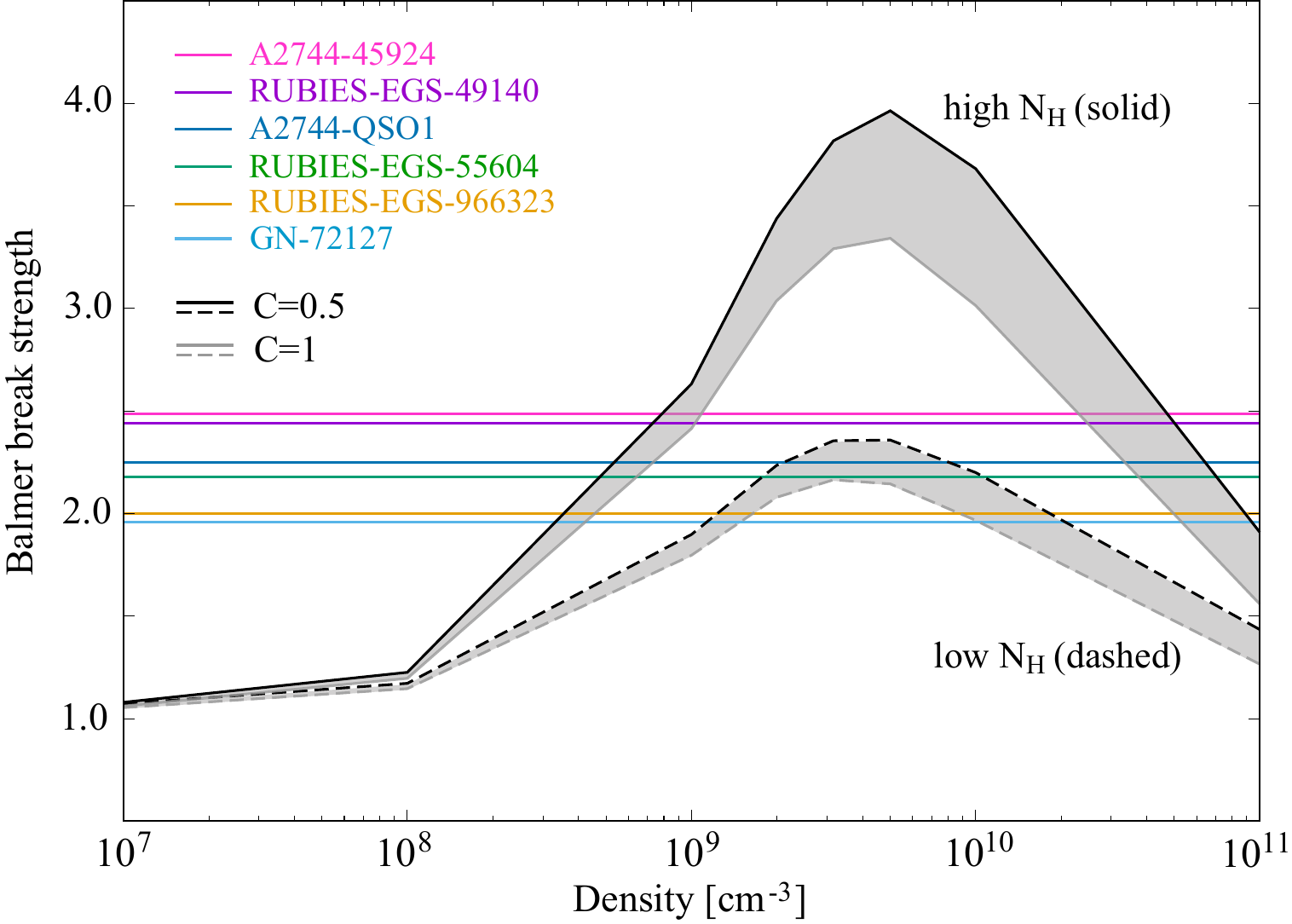}
\caption{The Balmer break strength defined by $F_{\lambda}(\lambda_{\rm B,red})/F_{\lambda}(\lambda_{\rm B,blue})$ 
for $N_{\rm H}=5.4\times 10^{22}~{\rm cm}^{-2}$ (dashed) and $1.7\times 10^{23}~{\rm cm}^{-2}$ (solid).
Two covering fractions are considered: $C=0.5$ (black) and $C=1$ (gray).
For the fiducial case (black and dashed curve), the Balmer break strength reaches values of $\geq 2$ in the density range of $10^9 \lesssim n_{\rm H}/\cc \lesssim 2\times 10^{10}$.
With increasing column density, the Balmer break becomes more prominent due to the enhanced column density of atomic hydrogen in the $n=2$ state.
These depths are consistent with those of six broad-line LRDs that show a Balmer break in the PRISM spectrum \citep[][see text]{Furtak_2024,Wang_2024b,Kokorev_2024b,Labbe_2024b}.
}
\label{fig:depth}
\vspace{3mm}
\end{figure}

Figure~\ref{fig:depth} presents the Balmer break strength of the total AGN SEDs as a function of slab density.
We measure the Balmer break strength using the ratio of the fluxes on the blue ($\lambda_{\rm B,blue}$) and red sides ($\lambda_{\rm B,red}$) of the Balmer limit.
To compare our result with LRDs that show both broad Balmer lines and a Balmer break in the spectra reported by \cite{Wang_2024b}, we adopt 
$\lambda_{\rm B,blue}=3600~{\rm \AA}$ and $\lambda_{\rm B,red}=4000~{\rm \AA}$.
We here show cases for two different column densities: $N_{\rm H}=5.4\times 10^{22}~{\rm cm}^{-2}$ (dashed; fiducial model) and $1.7\times 10^{23}~{\rm cm}^{-2}$ (solid).
These column densities are adjusted to maintain $A_V=3~{\rm mag}$, which thus corresponds to metallicities of $10^{-1}~\zsun$ and $10^{-1.5}~\zsun$, respectively.
For the fiducial case ($N_{\rm H}=5.4\times 10^{22}~{\rm cm}^{-2}$), the Balmer break strength reaches values of $\geq 2$ in the density range
of $10^9 \lesssim n_{\rm H}/\cc \lesssim 2\times 10^{10}$.
With increasing column density, the Balmer break becomes more prominent due to the enhanced column density of atomic hydrogen in the $n=2$ state.
The density range with a Balmer break strength $\geq 2$ is extended to $3\times 10^8 \lesssim n_{\rm H}/\cc \lesssim 10^{11}$ for the case of
$N_{\rm H}=1.7\times 10^{23}~{\rm cm}^{-2}$.
These trends hold even when the higher covering fraction is set ($C=1$; gray curves), though these cases show slightly weaker Balmer break depths 
compared to those with $C=0.5$ (black curves).
The measured strengths are consistent with observations of RUBIES-EGS-49140, 55604, and 966323 \citep{Wang_2024b}, A2744-QSO1 \citep{Furtak_2024}, 
GN-72127 \citep{Kokorev_2024b}, and A2744-45924 \citep{Labbe_2024b}.
At the high-density limit of $n_{\rm H}\gg 10^{10}~\cc$, the Balmer-break strength weakens due to the contributions of the Balmer jump from the nebular emission.

A sufficiently high hydrogen column density, likely attributed to dense absorbers such as clouds in the BLR or its surroundings, 
is required to imprint a Balmer break on the AGN SED.
While BLR clouds, or more generally clouds located within the conventional sublimation radius, are typically considered dust-free, 
large dust grains ($a\gtrsim 0.06~\mum$) in dense clouds with $n_{\rm H}\gtrsim 10^{9-10}~\cc$ may survive even at BLR scales ($\sim 2~R_{\rm BLR}$) 
due to the effective thermal energy loss from their surfaces \citep{Baskin_2018}.
These grains are thermally decoupled from the surrounding hot gas \citep[e.g.,][]{Tanaka_Omukai_2014} and maintain their temperature below the sublimation threshold 
(i.e., $T_{\rm dust}<T_{\rm sub}\ll T$) when 
\begin{equation}
\frac{n_{\rm H}}{10^{15}~\cc}<2.4 
\left(\frac{T_{\rm dust}}{10^3~\K}\right)^4\left(\frac{T}{10^4~\K}\right)^{-3/2},
\label{eq:ncrit}
\end{equation}
where the dust Planck-mean opacity is approximately constant for $100~\K \lesssim T_{\rm dust} \lesssim T_{\rm sub}$.
As a result, dust grains can survive in dense absorbers with $n_{\rm H}\lesssim 10^{9-11}~\cc$, where the AGN SED exhibits a prominent 
Balmer break feature under a high column density of $N_{\rm H} \sim 10^{23}~{\rm cm}^{-2}$.
Assuming the depletion factor of metals onto dust grains is comparable to the present-day Galactic value ($f_{\rm dep} \sim 0.5$), 
the metallicity needs to be as low as $Z \sim 10^{-1.3}~\zsun$ to maintain $A_V\sim 3~{\rm mag}$.
At higher metallicities ($Z\gtrsim 10^{-0.5}~\zsun$), unless the depletion factor is significantly lower, the visual extinction could become so large that the AGN emission 
is severely obscured ($A_V\gtrsim 20$).
In such cases, these systems would likely be observed as Type II AGNs.

Intriguingly, dense absorbers containing large-size grains ($a_{\rm min}\sim 0.06~\mum$) yield a gray extinction curve
at wavelengths shorter than $\lambda \simeq 2\pi a_{\rm min}\simeq 3800~{\rm \AA}$.
This characteristic may help explain the v-shaped SEDs observed in LRDs, which consist of a red optical continuum and a UV excess with
a turnover wavelength near the Balmer limit \citep[see discussion in][]{Li_LRD_2024}.
Further studies of dust sublimation and (re-)formation mechanisms in dense BLR clouds will be crucial to understanding the properties 
of high-redshift AGNs identified through JWST observations.

\begin{figure*}
\centering
\includegraphics[width=84mm]{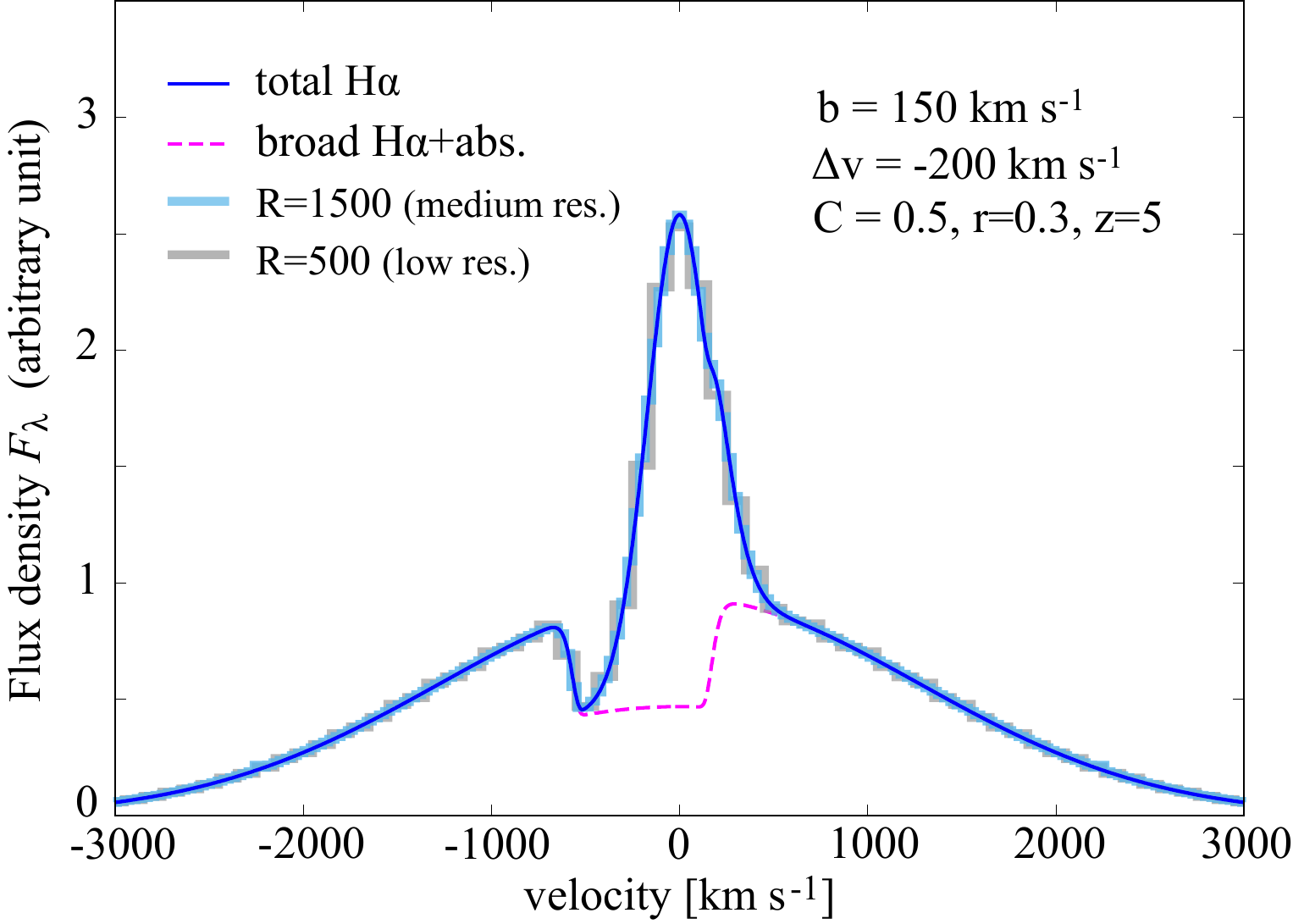} \hspace{3mm}
\includegraphics[width=84mm]{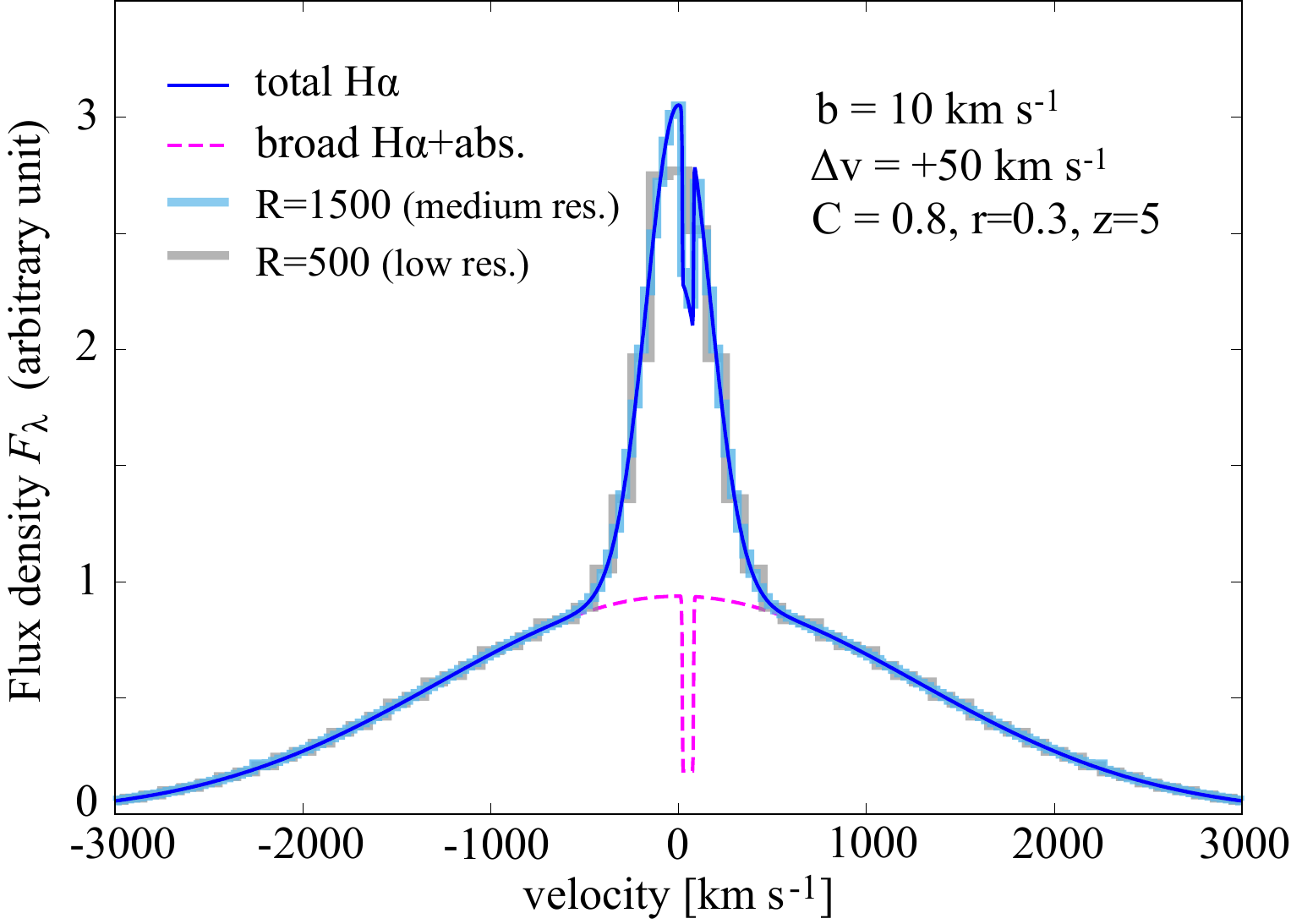}
\caption{
The H$\alpha$ line profiles with FWHM$_{\rm broad}=3000~\kms$, FWHM$_{\rm narrow}=400~\kms$ and $r=0.3$.
We explore two cases: $b=150~\kms$, $\Delta v=-200~\kms$, and $C=0.5$ (left panel) and $b=10~\kms$, $\Delta v=+50~\kms$, and $C=0.8$ (right panel).
The solid and dashed curves show the total H$\alpha$ line profile and the broad component with absorption by dense gas clumps with a column density of 
$n=2$ atomic hydrogen, $N_{{\rm H},n=2}=10^{16}~{\rm cm}^{-2}$. 
The total line profile with a finite spectral resolution ($R\equiv \Delta \lambda/\lambda_0 = 1500$ and $500$) is overlaid, where the source redshift is set to $z=5$.
}
\label{fig:line_profile}
\vspace{1mm}
\end{figure*}

\begin{figure*}
\centering
\includegraphics[width=84mm]{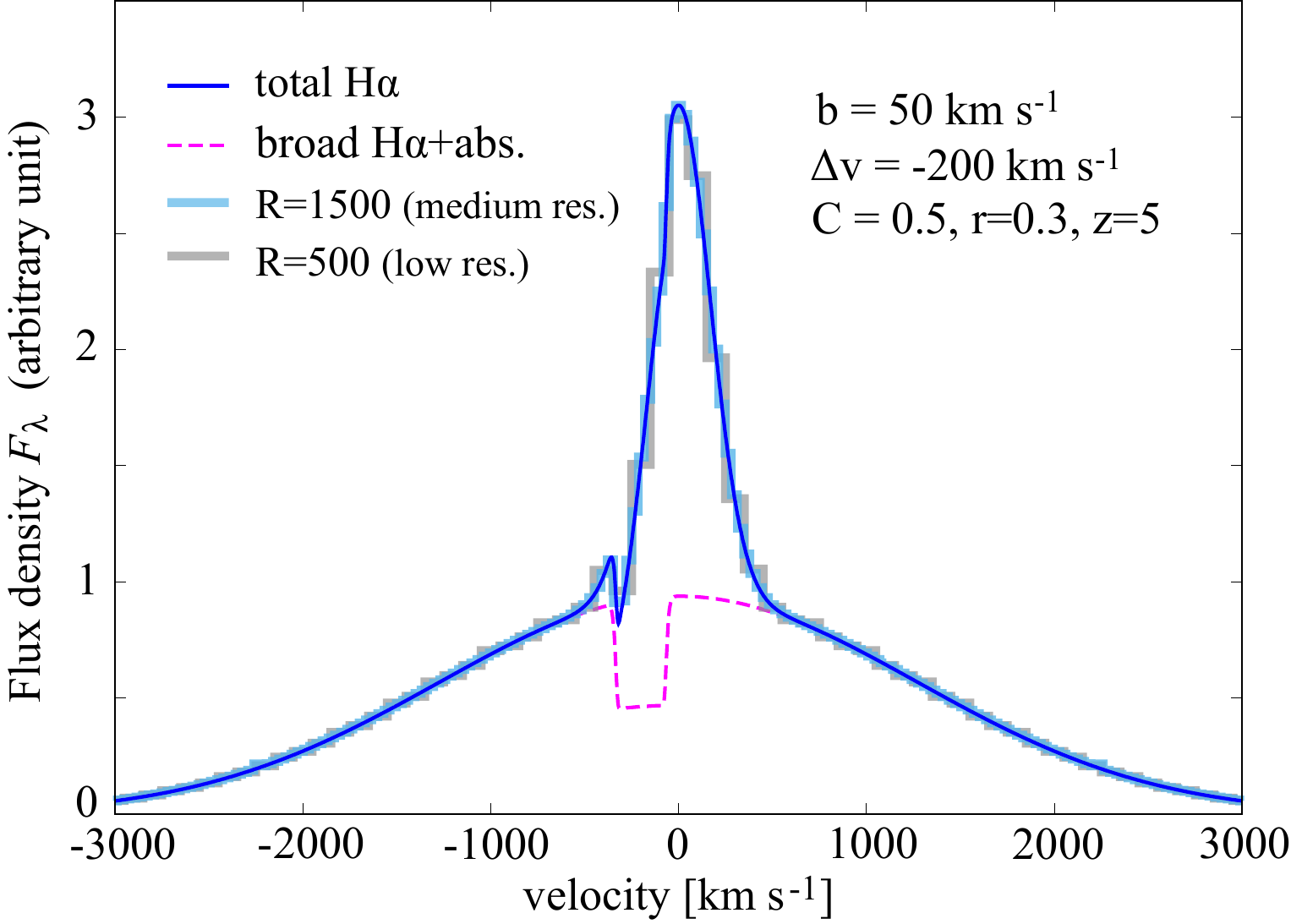}\hspace{5mm}
\includegraphics[width=84mm]{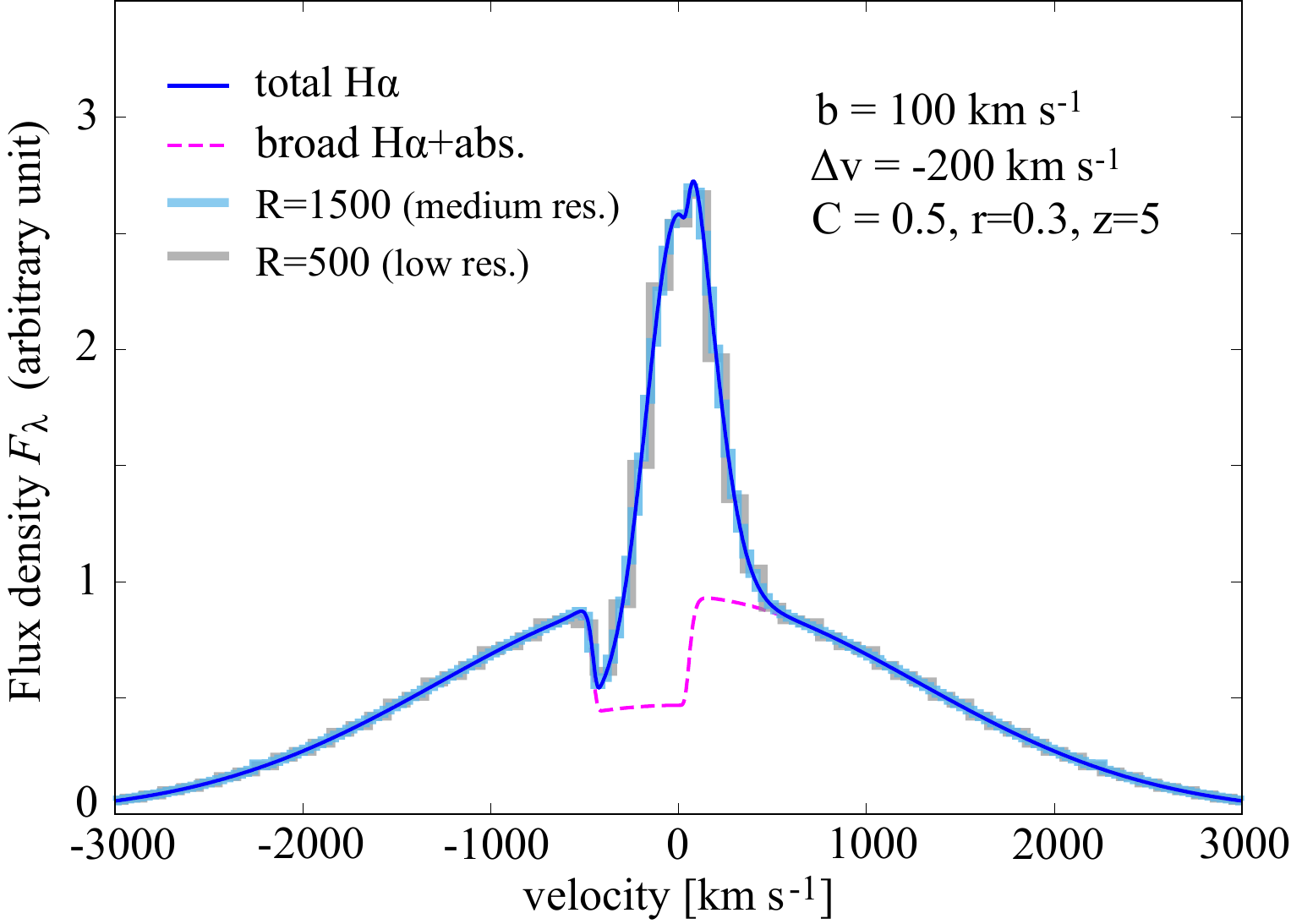}\\\vspace{3mm}
\includegraphics[width=84mm]{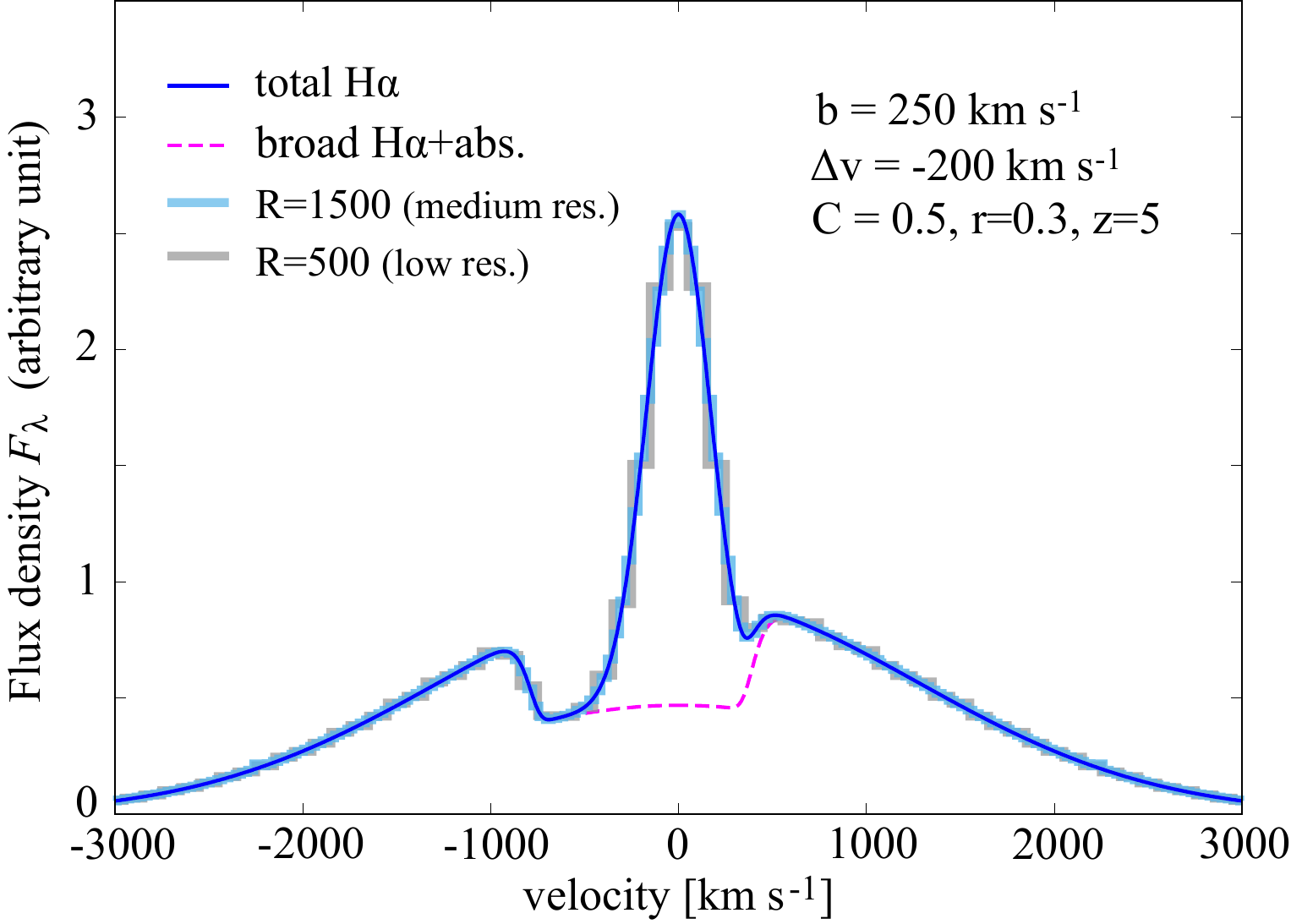}\hspace{5mm}
\includegraphics[width=84mm]{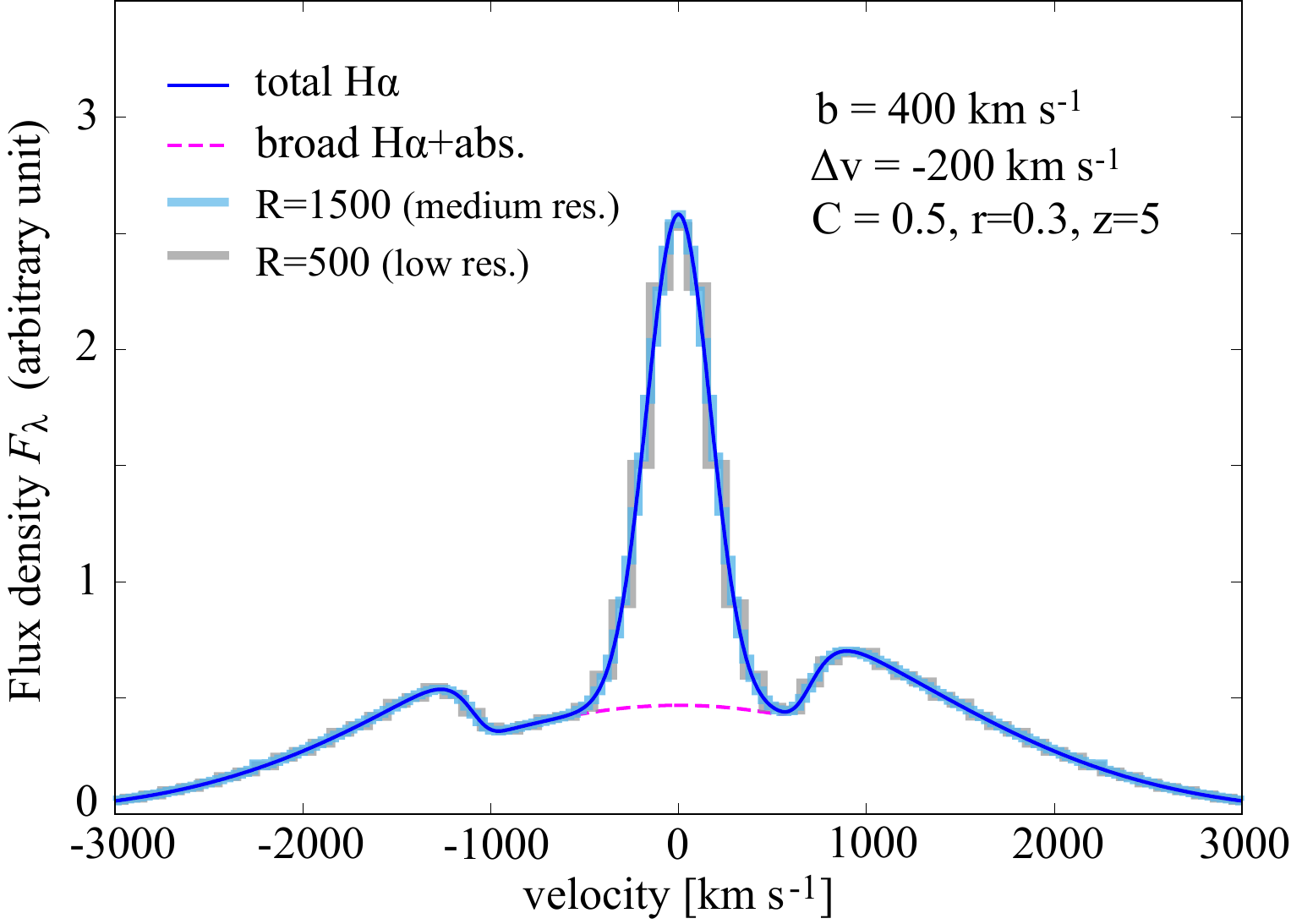}
\caption{
Same as in Figure~\ref{fig:line_profile}, but illustrating how the line shape changes with the width of the absorption feature, varying $b$ from $50~\kms$ to $400~\kms$.
The other parameters are identical to those used in Figure~\ref{fig:line_profile}.
}
\label{fig:line_profile_cases}
\vspace{1mm}
\end{figure*}

\section{Balmer absorption}\label{sec:absorption}

In this section, we examine how H$\alpha$ absorption modulates the line profile under the conditions that satisfy the criteria for producing a Balmer break, as discussed in Section~\ref{sec:break}.
The optical depth of a bound-bound transition between two (upper $u$ and lower $\ell$) energy states of atomic hydrogen is given by
\begin{align}
\tau_0 \approx \frac{\sqrt{\pi} e^2}{m_{\rm e}c} \frac{f_{\ell u}\lambda_{\ell u} N_{{\rm H}\ell}}{b}
= 314~N_{{\rm H}\ell,16} b_{200}^{-1},
\end{align}
\citep{Draine_2011}, where $e$ is the elementary charge, $m_{\rm e}$ the electron mass, $f_{\ell u}$ the oscillation strength between the two states, and $\lambda_{\ell u}$ the wavelength of a photon emitted in the transition.
For H$\alpha$ ($u=3$ and $\ell =2$), $f_{\ell u}=0.6047$ and $\lambda_{\ell u}=6563~{\rm \AA}(=\lambda_0)$ \citep[e.g.,][]{Drawin_1969}.
The optical depth at a frequency $\nu$ near the line center $\nu_0$ is approximated as $\tau_\nu =\tau_0 e^{-(u/b)^2}$, with $u=c(1-\nu/\nu_0)$ and $b=\sqrt{2}\sigma_v$,
where $\sigma_v$ is the one-dimensional velocity dispersion.
Following \cite{Juodzbalis_2024}, the absorption feature at a wavelength $\lambda$ is modeled using the attenuation formula,
\begin{equation}
f_\lambda = 1-C +C e^{-\tau_\lambda},
\end{equation}
where 
\begin{equation}
\tau_\lambda = \tau_0 \exp \left[-\frac{(\Delta v - c\lambda/\lambda_0)^2}{b^2}\right],
\end{equation}
with $\Delta v$ representing the velocity shift, where a negative value indicates a blueshift.

We apply the attenuation feature to the broad-line component of the H$\alpha$ emission, while excluding the narrow-line component.
This approach is based the idea that dense gas clumps are located between the BLRs and narrow-line regions (see also Section~\ref{sec:discussion1}), 
and thus absorb only the broad-line emission component. 
The H$\alpha$ line profile is modeled as
\begin{equation}
    F_\lambda = \phi_\lambda (\lambda_0, \sigma_{\rm broad})f_\lambda + r \phi_\lambda (\lambda_0, \sigma_{\rm narrow}),
\end{equation}
where $\phi_\lambda (\lambda_0, \sigma)$ is a Gaussian function with a mean $\lambda_0$ and dispersion $\sigma$.
The parameters $\sigma_{\rm broad}$ and $\sigma_{\rm narrow}$ represent the one-dimensional velocity dispersion for the broad and narrow components, respectively,
and $r$ is the relative ratio between the narrow and broad components before accounting for absorption.
In this analysis, we do not consider the continuum flux, for which the same level of attenuation should be applied. In this case, the apparent absorption feature becomes deeper as the unattenuated flux is higher, i.e., $\Delta F_\lambda \propto F_{\lambda, \rm broad+cont}$.

In Figure~\ref{fig:line_profile}, we present the H$\alpha$ line profile with FWHM$_{\rm broad}=3000~\kms$, FWHM$_{\rm narrow}=400~\kms$, and $r=0.3$.
The choice of the FWHM values aligns with the average ones for broad H$\alpha$ emission of LRD samples in \cite{Matthee_2024}.
We explore two cases: (1) $b=150~\kms$, $\Delta v=-200~\kms$, and $C=0.5$ and (2) $b=10~\kms$, $\Delta v=+50~\kms$, and $C=0.8$.
In the first case, the broad emission line shows a saturated, box-shaped absorption profile. 
Due to the blueshift of the absorption line center, the absorption appears just to the blue side of the line center when combined with the narrow emission line component. 
In the second case, the absorption depth is more profound because of the narrower absorption width and larger covering fraction. 
Since the absorption is slightly redshifted, the absorption feature are present on top of the narrow-line emission profile.
Those profiles resemble the spectral shapes seen in GOODS-N-9771 and J1148-18404, two LRD samples reported in \cite{Matthee_2024}, respectively.
To explain the line profile of J1148-18404 in our model, a substantially high covering fraction ($C\gtrsim 0.8$) is needed. Along with the slightly redshifted center ($\Delta v\simeq +50~\kms$),
the absorption may trace dense inflowing gas into the nucleus.

Figure~\ref{fig:line_profile_cases} illustrates how the spectral shape changes with different widths of the absorption feature, varying $b$ from $50~\kms$ to $400~\kms$.
For the case of $b=50~\kms$, the absorption feature is narrow and thus can be resolved with a spectral resolution of $R\gtrsim 1500$ (medium resolution).
As the width increases and approaches the velocity shift ($\Delta v=-200~\kms$), absorption begins to affect the redder side of the emission line.
In the most extreme case, where the absorption width is significantly large, the absorption profile becomes saturated and box-shaped troughs are imprinted on both blue and red side of the narrow-line emission.

\section{Discussion}\label{sec:discussion}

\subsection{Inflow, outflow, and BH feeding rates}
\label{sec:discussion1}

Using the properties of dense gas clumps measured from spectral line analyses, one can estimate the mass inflow, outflow, and BH feeding rate.
The detection of Balmer absorption blueshifted by $\Delta v \sim {\rm a~few}\times 100~\kms$ suggests the presence of dense neutral outflows in the nucleus.
Estimating the maximum outflow velocity as $v_{\rm out}= |\Delta v| + 2 \sigma_v \simeq 2 |\Delta v|$, the outflow condition yields $v_{\rm out}\gtrsim \sqrt{2GM_\bullet /R}$, 
where $R$ is the distance of the dense gas from the central BH with a mass of $M_\bullet$,
\begin{equation}
R \sim  \frac{GM_\bullet}{2(\Delta v)^2} \simeq 0.5~\pc ~M_7 \Delta v_{200}^{-2},
\end{equation}
where $\Delta v_{200}=\Delta v/(200~\kms)$ and $M_7=M_\bullet/(10^7~\msun)$.
We note that this distance estimate agrees to that derived from $\log U=-1.5$ (see Section~\ref{sec:break}), assuming $n_{\rm H}=3\times 10^9~\cc$ and the bolometric luminosity to be the Eddington value for $M_7=1$.

The mass outflow rate is calculated by
\begin{equation}
\dot{M}_{\rm out} = 4\pi C \mu m_{\rm p}N_{\rm H}Rv_{\rm out}
 \sim 1.2~\msunyr M_7 |\Delta v_{200}|^{-1},
 \label{eq:outflow}
\end{equation}
where $\mu=1.22$ is the mean molecular weight, and $N_{\rm H}=5\times 10^{22}~{\rm cm}^{-2}$ and $C=1$ are set\footnote{
We consider a continuous density distribution with some degree of clumpiness 
in the outflowing region, rather than a single thin-shell structure. Such profiles arise from continuous mass loading by disk winds,
as predicted by numerical simulations \citep{Ohsuga_2009, Yuan_Narayan_2014,Hu_2022a}.
Regardless of the origin of the cloud clumpiness, the column density can be determined as $N_{\rm H}\simeq n_{\rm H}(R)R$. 
Given a column density, the degree of clumpiness can be adjusted by increasing the density $n_{\rm H}$ while decreasing 
the slab thickness $\Delta s$ (see Section~\ref{sec:break}).}.
This outflow rate is $\simeq 5$ times higher than the Eddington accretion rate, $\dot{M}_{\rm Edd}\equiv 0.23 M_7~\msunyr$.
The ratio of $\dot{m}_{\rm out}\equiv \dot{M}_{\rm out}/\dot{M}_{\rm Edd}\simeq 5$ is independent of the BH mass.
In a steady state, mass conservation gives the BH feeding rate as $\dot{M}_\bullet = \dot{M}_{\rm in}-\dot{M}_{\rm out}(\geq 0)$.
When the gas supplying rate from larger radii exceeds the Eddington rate, i.e., $\dot{m}_{\rm in}\equiv \dot{M}_{\rm in}/\dot{M}_{\rm Edd}>1$, 
radiation-driven outflows carry the inflowing mass away and decreases the BH feeding rate from the original inflow rate \citep[e.g.,][]{Jiang_2014,Yuan_Narayan_2014}.
Adopting a mechanical feedback model obtained in radiation hydrodynamic simulations of gas accretion at a vicinity of a BH, the BH feeding rate 
is given by a scaling relationship of $\dot{m}_\bullet \simeq \dot{m}_{\rm in}^{1/2}$ \citep{Hu_2022a}.
Given the formula, the inflow and BH feeding rates can be derived as 
\begin{align}
\dot{m}_{\rm in} &= \frac{1+2\dot{m}_{\rm out}+\sqrt{1+4\dot{m}_{\rm out}}}{2}\simeq 7.8,\nonumber\\
\dot{m}_\bullet  &= \frac{1+\sqrt{1+4\dot{m}_{\rm out}}}{2}\simeq 2.8,
\end{align}
for $\dot{m}_{\rm out}\simeq 5$.
Thus, the BH in this system grows in mass at a moderately super-Eddington rate.

\subsection{Radiative signatures of rapidly growing BHs embedded in dense environments}
\label{sec:discussion2}

We have shown that dense circum-nuclear gas can produce a Balmer break in the AGN continuum spectrum and cause absorption in the H$\alpha$ and H$\beta$ emission line.
Here, we discuss additional features of the spectrum under these circumstances. 
In conditions of $n_{\rm H}\gtrsim 10^{9-10}~\cc$, where H$\alpha$ absorption ($n=2 \rightarrow 3$) arises from hydrogen gas in the first-excited $n=2$ states, 
the Ly$\beta$-line transition ($n=3 \rightarrow 1$) also occurs frequently but these photons are effectively trapped within such a dense medium.
Resonance fluorescence by Ly$\beta$ leads to excitation of neutral oxygen and produces three \ion{O}{1} emission lines ($\lambda$1304, $\lambda$8446, and $\lambda$11287), 
owing to a coincidence of energy levels between neutral oxygen and hydrogen \citep{Kwan_Krolik_1981}.
Indeed, these \ion{O}{1} lines are observed in low-redshift AGNs as a proxy of dense circum-nuclear regions \citep[e.g.,][]{Grandi_1980,Martinez-Aldama_2015,Cracco_2016}
and show good correlations with the properties of other low-ionization lines, such as \ion{Ca}{2} and \ion{Fe}{2}, which originate from the same portion of BLR clouds 
\citep{Rodriguez-Ardila_2002a,Riffel_2006,Matsuoka_2007,Matsuoka_2008}.
The usefulness of these \ion{O}{1} emission lines have been noted in \cite{Inayoshi_2022b}, who studied the accretion process of seed BHs in early, metal-poor protogalaxies 
by performing radiation hydrodynamic simulations.
Their work has claimed that JWST NIRSpec observations of low-ionization \ion{O}{1} emission lines can test whether the BH is fed via a dense accretion disk at super-Eddington rates.

Intriguingly, two of the \ion{O}{1} emission lines ($\lambda$8446, $\lambda$11287) have been detected in some JWST-identified AGNs;
GN-28074 at $z_{\rm spec}=2.26$ \citep{Juodzbalis_2024} and RUBIES-BLAGN-1 at $z_{\rm spec}=3.1$ \citep{Wang_2024a}\footnote{
GN-28074 shows clear blueshifted absorption features on the H$\alpha$, H$\beta$, and \ion{He}{1}~$\lambda 10830$ emission lines \citep{Juodzbalis_2024}.
RUBIES-BLAGN-1 exhibits a strong blueshifted absorption feature in \ion{He}{1}~$\lambda 10830$, within a wavelength coverage of the G395M spectra. 
However, The presence of the Balmer absorption in this object remains unclear because of the lower spectral resolution of the PRISM data \citep{Wang_2024a}.
}.
For GN-28074, the flux ratio of the \ion{O}{1} emission lines is close to unity, indicating that they are likely produced by Ly$\beta$ fluorescense.
However, constraints on \ion{O}{1}~$\lambda$1304 line are required to establish the Ly$\beta$ pumping scenario in a robust way 
\citep{Rodriguez-Ardila_2002b}, while this line is not covered by the wavelength range of the NIRSpec spectrum of GN-28074.
Despite the difficulty of confirmation, the Ly$\beta$ fluorescense scenario is consistent with the presence of dense gas optically thick to Balmer lines 
and photons with shorter wavelengths from the Balmer limit.
Similarly, \citet{Wang_2024a} also reported the spectrum of an AGN that exhibits both \ion{O}{1}~$\lambda$8446 and $\lambda$11287 lines,
though the detailed analysis has not been performed for the \ion{O}{1} lines.

Another radiative signature of rapidly accreting BHs is a prominent H$\alpha$ emission line with a large equivalent width (EW), which is predicted to be EW$_{\rm H\alpha,0}\simeq 500-1500~{\rm \AA}$ due to efficient collisional excitation to $n=3$ states in a dense super-Eddington accretion disk \citep{Inayoshi_2022b}.
This prediction aligns with observations of JWST-identified AGNs at $z > 4$, which typically show EWs approximately three times higher than those of quasars at lower redshifts of $z < 0.6$ \citep{Maiolino_2024b}.
The higher EW suggests that the JWST AGN population is embedded by dense gas with a high covering fraction, enhancing the reprocess efficiency of H$\alpha$ emission.
In addition, super-Eddington accreting (seed) BHs at $z \gtrsim 8$ are expected to be detectable through a unique color excess in the JWST NIRCam/MIRI bands the redshifted 
H$\alpha$ line enters \citep{Inayoshi_2022b}.
A larger samples of broad-H$\alpha$ emitters will bring insights on the properties of JWST-identified AGNs (e.g., H$\alpha$ EWs, Eddington ratios, and BH masses; see a recent work by \citealt{Lin_2024}).

\subsection{Steepness and depth of a Balmer break}

In our model, where JWST AGNs are embedded in very dense gas clumps, the Balmer break-like spectral features observed in 
some LRDs are attributed to absorption at wavelengths shorter than the Balmer limit by these dense gas clouds. 
However, the discontinuity in the spectrum near the Balmer limit, as shown in Figure~\ref{fig:break}, appears more abrupt 
than what is observed in the PRISM data for these LRDs. 
This difference arises because our model employs a simplified absorber with a uniform density and non-turbulent slab, leading to a nearly isothermal 
temperature structure ($T\simeq 8000-10^4~\K$).
For comparison, individual stellar spectra with surface temperatures of $\simeq 8000-10^4~\K$ show a profound and steep Balmer break \citep{Kurucz_1979,Poggianti_1997}. 
In contrast, spectra near the Balmer limit in cooler, lower-mass stars are substantially smoother, creating a gradual Balmer break in galaxies 
where long-lived, low-mass stars dominate the light \citep[e.g.,][]{Worthey_1994}. 
By this analogy, a smooth Balmer break could also be produced in our model if the density structure of absorbers were non-uniform, turbulent, and multi-phased \citep[e.g.,][]{Wada_2016}, including cooler region.
Indeed, \citet{Ji_2025} find that microscopic turbulence makes the Balmer break feature smoother, better matching the spectral shapes observed in some LRDs.
While a detailed analysis of these effects is beyond the scope of our work, future observations that measure the continuum shapes of JWST AGNs in larger samples 
could potentially constrain the density and temperature structure of the absorbing gas.

Moreover, the depth of the Balmer break in our model depends on the hydrogen column density of gas absorbers.
For the six LRDs shown in Figure~\ref{fig:depth}, the narrow range of Balmer break strengths ($\sim 2-2.5$) indicates the need for a specific column density.
However, this narrow range might also reflect a photometric selection bias, as LRDs are typically identified with a consistent reddening level ($A_V \simeq 3~{\rm mag}$).
The column density conditions naturally produce Balmer break strengths within the observed range ($\simeq 2-3$).

In contrast, the stellar-origin scenario explains the Balmer break feature through stellar populations of specific ages (several hundred Myr; see Figure A1 in \citealt{Wang_2024b}).
Although these models predict an upper limit for the Balmer break strength of $\lesssim 2.5$, marginally consistent with the brightest LRD (A2744-45924; \citealt{Labbe_2024b}),
fine-tuning of stellar ages and populations are required to match the observations.
To differentiate these scenarios, discovering a deeper Balmer break feature that cannot be reproduced by stellar-origin models would provide decisive evidence supporting
the AGN-origin scenario.

\subsection{Implications of stellar populations in LRDs
and ultra-massive quiescent galaxies at high redshifts}

Thus far, we have observed several LRDs that show continuum spectra with a Balmer break.
If these LRDs are powered by starbursts alone, an extremely massive stellar mass comparable to that of the Milky Way ($M_\star \sim 10^{11}~\msun$) would be derived \citep{Furtak_2024,Wang_2024b,Baggen_2024}.
Considering their high abundance within a cosmic volume ($\sim 10^{-5}~\mpc^{-3}$), the inferred stellar mass density would exceed the theoretical upper bound in a flat $\Lambda$ cold-dark-matter (CDM) universe, with a $100\%$ baryon-to-star conversion factor or $\gtrsim 10-20\%$ \citep{Wang_2024b,Akins_2024}. 
Additionally, in this stellar-origin hypothesis, the very compact nature of LRDs suggests the presence of unrealistically dense stellar clusters in these LRDs, with surface density above $\gtrsim 10^6~\msun~\pc^{-2}$ \citep[e.g.,][]{Baggen_2024}, which is one order of magnitude higher than the densest star clusters or the densest elliptical galaxy progenitors \citep{Hopkins_2010,Baggen_2023}.
If this scenario is true, a large number ($\gg 10~{\rm yr}^{-1}$) of tidal disruption events (TDEs) would be observed even within a small area ($\lesssim 0.1~{\rm deg}^2$) of deep JWST surveys \citep{Inayoshi_2024}\footnote{\cite{Inayoshi_2024} estimated the stellar mass density as $\sim 5\times 10^4~\msun~\pc^{-2}$, based on the dust mass required to achieve an extinction level of $A_V\simeq 3$ mag, which explains the red continua observed in LRDs. Thus, the predicted TDE detection number is as low as $\sim 2-10~(0.2-2)~{\rm yr}^{-1}$ for the JADES-Medium (and COSMOS-Web, respectively).}.

In contrast, under the scenario where the Balmer break is caused by gas absorption surrounding the AGN, the stellar mass inferred from SED fitting can be dramatically reduced, thereby resolving the tension with the $\Lambda$CDM framework. This interpretation also aligns with the observation that 
LRD are extremely compact, possibly dominated by an unresolved source in most cases.
Moreover, this supports an idea that the BH population at the early epochs tends to be overmassive relative to the mass correlation with their host mass observed in the local universe \citep[e.g.,][]{Maiolino_2023_JADES,Harikane_2023_agn},
as suggested by numerical simulations that sufficiently resolve the galactic nuclear scales at $\lesssim~0.1-1~\pc$ \citep[e.g.,][]{Inayoshi_2022a,Inayoshi_2022b}.

Additionally, recent JWST observations have revealed very massive quiescent galaxies in the distant universe that have already quenched star formation at $z>6$ \citep{Carnall_2023,Carnall_2024,deGraaff_2024,Weibel_2024}. Some of these galaxies also show broad H$\alpha$ emission, indicating the presence of AGNs (and in others the presence of an AGN cannot be easily excluded). Similar to LRDs, these massive galaxies face the same issue of exceeding the $\Lambda$CDM stellar-mass density limit. 
However, if the Balmer break and absorption features in their spectra are partly due to dense gas absorption, and their continua are contributed by an AGN, this could help alleviate some of the inferred cosmological tensions.

\subsection{\ion{He}{1} absorption and emission}

Similar to Balmer lines, a blueshifted absorption feature is also observed in the \ion{He}{1}~$\lambda 10830$ emission line for the two JWST-identified AGNs, which simultaneously show Balmer absorption \citep[][see discussion in Section~\ref{sec:discussion2}]{Wang_2024a,Juodzbalis_2024}.
For the two objects, the velocity shift tends to be larger than that of the Balmer lines.
This absorption is caused by the metastable triplet state $2 ^3S$ (or $1s2s~ ^3S_1$) for helium with a column density of $N_{\rm He}(2 ^3S)\sim (1-3)\times 10^{14}~{\rm cm}^{-2}$.
Based on photoionization models, \ion{He}{1} absorption features consistent with the observations can be reproduced under the conditions where Balmer absorption becomes prominent \citep{Juodzbalis_2024}.

We also note that collisional excitation from the metastable $2 ^3S$ state to higher excitation levels can enhance the intensity of three emission lines of \ion{He}{1}~$\lambda 3890$, $\lambda 5877$, and $\lambda 7067$ (corresponding to transitions from the $1s3s$, $1s3p$, $1s3d$ states of triplet helium to the lower energy states; see Figure~14.3 and Table 14.5 of \citealt{Draine_2011}). The latter two emission lines are {\it indeed} observed in the spectra of the two sources mentioned above, although detailed analyses have not been conducted for the \ion{He}{1} emission lines (see Figure~2 of \citealt{Wang_2024a} and Figure~1 of \citealt{Juodzbalis_2024}).
Further explorations of the extremely dense interstellar medium within high-redshift AGNs, particularly those identified through JWST observations, will provide deeper insights into the physics of early BH assembly.

\newpage

\acknowledgments
We greatly thank Yuhiko Aoyama, Seiji Fujimoto, Jenny Greene, Kevin Hainline, Jakob Helton, Luis C. Ho, Harley Katz, Kei Tanaka, and Bingjie Wang for constructive discussions.
K.~I. acknowledges support from the National Natural Science Foundation of China (12073003, 11721303, 11991052), 
and the China Manned Space Project (CMS-CSST-2021-A04 and CMS-CSST-2021-A06). 
R.~M. acknowledges support by the UKRI Frontier Research grant RISEandFALL and from a research professorship from the Royal Society.
This research was supported in part by grant NSF PHY-2309135 to the Kavli Institute for Theoretical Physics (KITP). 
We are deeply grateful to Raffaella Schneider, Rachel Somerville, Brant Robertson, and Volker Bromm for organizing the KITP workshop, 
``Cosmic Origins: The First Billion Years", which provided the initial inspiration for this research.

\bibliographystyle{aasjournal}
\bibliography{ref.bib}


\end{document}